# Synchronization of spin-torque driven nanooscillators for point contacts on a quasi-1D nanowire: Micromagnetic simulations

D.V. Berkov*

*Innovent Technology Development, Pruessingstr. 27B, D-07745, Jena, Germany*

## Abstract

In this paper we present detailed numerical simulation studies on the synchronization of two spin-torque nanooscillators (STNO) in the quasi-1D geometry: magnetization oscillations are induced in a thin NiFe nanostripe by a spin polarized current injected via square-shaped CoFe nanomagnets on the top of this stripe. In a sufficiently large out-of-plane field, a propagating oscillation mode appears in such a system. Due to the absence of the geometrically caused wave decay in 1D systems, this mode is expected to enable a long-distance synchronization between STNOs. Indeed, our simulations predict that synchronization of two STNOs on a nanowire is possible up to the intercontact distance $\Delta L = 3$ mkm (for the nanowire width $w = 50$ nm). However, we have also found several qualitatively new features of the synchronization behaviour for this system, which make the achievement of a stable synchronization in this geometry to a highly non-trivial task. In particular, there exist a *minimal* distance between the nanocontacts, below which a synchronization of STNOs can not be achieved. Further, when the current value in the first contact is kept constant, the amplitude of synchronized oscillations depends *non-monotonously* on the current value in the second contact. Finally, for one and the same currents values through the contacts there might exist several synchronized states (with different frequencies), depending on the initial conditions.

___________________

*e-mail: db@innovent-jena.de

# I. INTRODUCTION

Since the theoretical prediction [1, 2] and first reliable experimental confirmations [3, 4] of the possibility to excite magnetization dynamics in thin ferromagnetic films using a spin-polarized current (spin-torque), an enormous effort was devoted to a thorough fundamental understanding and development of various practical applications of this novel physical effect (see, e.g., the reviews [5, 6, 7, 8, 9, 10, 11] and references therein).

Spatial attention has been paid to the spin-torque induced generation of a microwave signal in various geometries. First of all, this process is highly interesting from the fundamental point of view, being an almost unique example of strongly non-linear oscillations in magnetic systems (i.e., oscillations with a strong frequency dependence on the oscillation amplitude) [10, 12, 13]. This non-linearity results in several qualitatively important dynamical effects, like the appearance of bullet-like localized oscillation modes [14], opposite frequency shifts with increasing external field for in-plane and out-of-plane field orientations [10, 13], unusually large bandwidth interval for the phase-locking of corresponding oscillators [15] etc. Second, nanoscale microwave emitting systems are of a large practical interest, allowing in principle the creation of generators with the sizes less than 1 mkm for the wide frequency regions (from several tens of MHz to several hundreds of GHz), which frequency could be tuned simply by changing the dc current strength. For this reasons a very large effort has been applied for experimental studies of these microwave oscillators, with the aim to the decrease the oscillator linewidth and to achieve the largest possible microwave emission power [16, 17, 18, 19, 20, 21, 22]. Although a substantial progress could be achieved in both directions, corresponding parameters of a single oscillator are still not good enough for potential applications (including detection of radio frequency magnetic fields and telecommunication).

Decisive improvement of the characteristics of such devices could be obtained when oscillation of several such STNOs are synchronized (phase-locked), leading both to a large increase of the output oscillation power (~ $N^2$ for $N$ synchronized oscillators) and significant narrowing of the generated linewidth. During the last years, an impressive deepening of our theoretical understanding of the mutual synchronization of two and more STNOs has been achieved [15, 23, 24, 25, 26]. Even a rather complicated system of two dipolarly coupled STNOs with the vortex magnetization configuration was studied analytically and numerically [27]. Unfortunately, experimental progress on this topic remains rather moderate: after two pioneering reports on the synchronization of two STNOs implemented as closely placed nanocontacts on a thin magnetic film [28, 29], only in one paper [30] the synchronization of four nanooscillators – arranged as a matrix 2 x 2 – has been demonstrated experimentally.

The major difficulties arising by the experimental realization of the STNO synchronization are not only due to very high demands on the device quality, where the accurate patterning of nanolayer stacks with the spatial resolution of several nanometers is required. An additional problem is the relatively fast decay of the spin wave amplitude (~ $1/r$) emitted out of the point contact area in the 2D geometry used up to now in all synchronization experiments. Taking into account, that the spin wave propagation is the major mechanism of STNOs coupling for typical distances between the nanocontacts (~ 1 mkm) [23], this geometrically caused wave decay leads to a very small maximal intercontact distance $L_{max} \approx 500$ nm, for which the synchronization still can be obtained [28, 29, 30].

This insight led to the natural idea to use for the synchronization of STNOs a quasi-1D system, i.e., a thin nanostripe with a small width about 50 - 100 nm, with point contacts placed on the top of this stripe (we note in passing, that such devices have been used to verify the analytical theory of the STNO [31] and to study the non-linear damping of spin waves using the spin torque ferromagnetic resonance [32]). A spin wave, emitted from one such point contact, would propagate along a quasi-1D stripe without the decay caused by the system dimensionality. This means that its amplitude would decrease only due to a 'normal' Gilbert damping of the magnetization oscillations. First attempts to find the synchronization in such a multicontact system applying an in-plane external field did not succeed [33], because, as we could show in our detailed numerical study of this system [34], all its oscillation modes in the in-plane field are localized under the point contact area, similar to the non-linear bullet in the 2D system [14].

However, in a sufficiently large *out-of-plane* external field a propagating mode was found in our simulations [34], so that this geometry (quasi-1D planar waveguide in an out-of-plane field) could be a promising candidate for the STNO synchronization. For this reason, in the present paper we perform a systematic numerical study of the STNO synchronization in the system consisting of two square-shaped point contacts placed on the top of a 50 nm wide nanowire. Our simulation methodology and system under study are described in detail in Sec. II. The influence of the 2[nd] point contact not carrying any current on the magnetization dynamics is explained in Sec. III.A. An overview of the synchronization behaviour in dependence on the intercontact distance is given in Sec. III.B, and the detailed consideration of especially important issues - the existence of multiply synchronized states and the non-monotonic power dependence on the injected current – is presented in Sec. III.C and III.D, respectively. We summarize our findings in Sec. IV.

## II. SIMULATED SYSTEM AND SIMULATION METHODOLOGY

To study the synchronization of current-induced magnetization oscillations in a quasi-1D two-contact device, we have simulated the system consisting of a NiFe (also Permalloy or Py) nanostripe having a rectangular cross-section with the thickness $h_{Py}$ = 6 nm and width w = 50 nm. Two square-shaped CoFe nanoelements with equal lateral sizes $50 \times 50$ nm$^2$ and the thicknesses $h_{CoFe}$ = 15 nm were placed on the top of this stripe. The interlayer spacer thickness between Py and CoFe was set to $h_{sp}$ = 8 nm. The system was discretized in-plane using lateral cell sizes 3.125 x 3.125 nm$^2$, so that CoFe squares were subdivided in-plane into 16 x 16 cells. No additional discretization was performed in the direction perpendicular to the stripe plane, so that the discretization cell size in this direction was 6 nm for Py and 15 nm for CoFe layers. The distance between the contacts $\Delta L$ (see the top of Fig. 2 for the definition of $\Delta L$) have been varied between 500 nm and 3000 nm to find out the region of intercontact distances where the synchronization can take place. Simulations were performed using the open boundary conditions. The length of the nanowire was always taken to be 1000 nm larger than the intercontact distance to ensure that the stray field from the nanowire ends does not affect the magnetization dynamics under the point contacts.

Magnetic parameters for the nanostripe material have been chosen similar to those used in [34]: magnetizations $M_{Py}$ = 580 emu/cm$^3$ and $M_{CoFe}$ = 1800 emu/cm$^3$ (standard value for the Co$_{50}$Fe$_{50}$ alloy) and exchange constants $A_{Py} = 0.65 \times 10^{-6}$ erg/cm and $A_{CoFe} = 3 \times 10^{-6}$ erg/cm. We remind, that chosen magnetization and exchange stiffness of the Py nanostripe provide the best fit to the experimentally measured ST-induced oscillation frequencies as explained in details in [34]. Physically these reduced values (compared to standard magnetic parameters of Py) are usually attributed to Cu diffusion into the thin Py stripe.

Simulations were carried out with our 35 software package [35], which dynamical part solves the Landau-Lifshitz-Gilbert equation of motion for magnetic moments. The damping in nanostripe and point contact materials was set to $\lambda$ = 0.01. In addition, increased damping towards the ends of the wire was assumed in order to avoid the artificial spin wave reflections from these ends (see [36] and [37] for details).

To include the spin torque, we have added to the effective field the term $\mathbf{H}_{ST} = f_J \cdot [\mathbf{M} \times \mathbf{p}]$ (it corresponds to the addition of the standard Slonczewski term $\mathbf{\Gamma} = -f_J(\theta) \cdot [\mathbf{M} \times [\mathbf{M} \times \mathbf{p}]]$ to the equation of motion in the Gilbert form). The ST-term depends on the angle $\theta$ between magnetizations of the 'soft' and 'hard' layers as $f_J(\theta) = a_J \cdot 2\Lambda^2 / ((\Lambda^2 + 1) + (\Lambda^2 - 1)\cos\theta)$ [34, 38]. Here $\Lambda^2 = 1 + \chi$ is related to the asymmetry parameter $\chi$ in the expression $\Delta R_{GMR} \sim (1 - \cos^2(\theta/2)) / (1 + \chi \cos^2(\theta/2))$ for the angular dependence of the GMR resistance. In our simulations we have set $\chi = 1$ [34, 38]. The spin polarization degree of the electric current entering the factor $a_J$ in the ST-term $f_J$ was chosen to be $P = 0.3$.

According to our previous results [34], propagating mode is observed in this system only in strong out-of-plane magnetic fields. For this reason, we have simulated the system dynamics in the constant external field $H_0 = 10$ kOe, directed almost perpendicular to the nanostripe plane.

For each system size and current values through the point contacts simulations have been performed for the 'physical' time duration $t_{max} = 200$ ns, starting from the time moment when the current(s) was (were) switched on. Oscillation power spectra displayed in next sections have been calculated using the time dependencies of the magnetization components after the steady-state precession regime was achieved. All spectra and the total power values shown in subsequent figures were computed using the magnetization values averaged only over the nanostripe regions under the point contacts.

To study whether different methods of the current increase affect the dynamical system behaviour, we have used two simulation protocols. In the protocol A the currents through both contacts were increased instantly from zero to their final values. In the protocol B, the current through the first contact $I_1$ was increased instantly, whereas $I_2$ was increased from zero to its final value *linearly* in time within $\Delta t = 10$ ns. As it will be shown below, these two protocols may lead to substantially different results, thus demonstrating the existence of multiply synchronized states for the same geometric and magnetic system parameters and current values.

### III. RESULTS AND DISCUSSION

### A. Effect of a neighboring contact not carrying any current

In order to fully understand the influence of the 2$^{nd}$ contact on the ST-induced magnetization dynamics in our case, we have first performed comparative modeling of two systems: the first system contained only one point contact, whereas in the second system both point contacts were present, but the current through the second contact was set to zero ($I_2 = 0$). The distance between the contacts in the latter case is $\Delta L = 460$ nm.

Comparison of these systems is presented in Fig. 1, where we show the oscillation frequency and power of the $m_x$-component (in-plane magnetization component along the stripe length) as functions of the current density $j$, for the current region where a propagating mode is observed. This comparison demonstrates that for the same current, oscillation *frequencies* for systems with one and two contacts are nearly equal (at least for currents where an appreciable oscillation power is found), whereas the oscillation *power* for the system with a single contact is 2 to 3 times larger than for the system with two contacts (we remind that in the latter case the current through the 2$^{nd}$ contact is zero). In addition, considerable power oscillations with the current strength are observed in the two-contacts system.

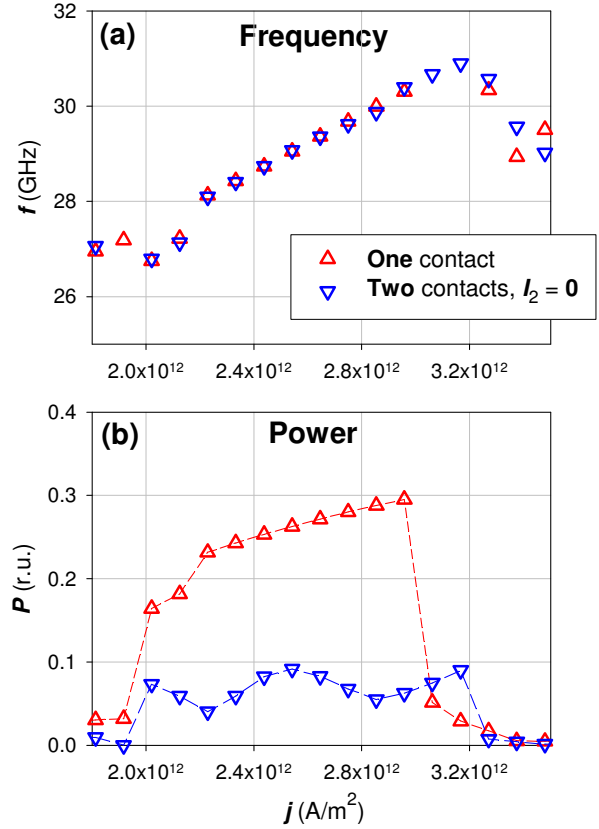

Fig. 1. (a) oscillation frequency $f$ and (b) oscillation power of the $m_x$-component $P_x$ (b) as functions of the current strength through the 1$^{st}$ point contact for two cases: (*i*) the 2$^{nd}$ contact is absent (red triangles up) and (*ii*) the 2$^{nd}$ contact is present, but does not carry any current (blue triangles down). The intercontact distance for the 2$^{nd}$ case is $\Delta L = 460$ nm.

These results can be understood in view of our findings concerning the dependence of the oscillation power on the magnetization value of the CoFe nanoelement in a similar system containing one contact only [34]. Namely, we have shown (see Fig. 7b and 8 in [34]), that a considerable amount of the energy pumped into the system via the fed of a *dc*-current may be spent for the oscillations of magnetization of a CoFe element. These oscillations are induced *not* by a spin torque acting on the CoFe magnetization by electrons polarized by a Py nanostripe (in our simulations we neglect this spin torque, because the thickness and magnetization of the CoFe layer are much larger than those of the Py stripe), but by a magnetodipolar interaction between magnetic moments of the 'soft' and 'hard' layers. In the system studied here, a similar process occurs: The (non-linear) propagating spin wave emitted from the 1$^{st}$ point contact - which amplitude decays relatively slowly due to the quais-1D nature of the Py waveguide - reaches the area under the 2$^{nd}$ contact and induces the magnetization oscillations of the 2$^{nd}$ CoFe nanoelement. As follows from our results presented in [34] and the comparison of oscillation powers shown in Fig. 1b, a substantial amount of energy can be spent to sustain these oscillations, thus decreasing the amplitude of the magnetization oscillations within the 1$^{st}$ contact area in the Py nanowire.

Oscillations of the power as the function of the current strength for the two-contacts system with $I_2 = 0$ can be also explained by the magnetodipolar interaction between the 2$^{nd}$ CoFe nanoelement and the Py nanostripe. Namely, a relatively strong magnetodipolar field from the CoFe nanorectangle causes an inhomogeneity of the total field and hence - of the magnetization configuration of a Py layer under the 2$^{nd}$ contact. Spin waves emitted from the 1$^{st}$ contact are partially reflected from this inhomogeneity, and the interference between the emitted and the reflected waves changes the amplitude of magnetization oscillations under the 1st contact. This interference is either constructive or destructive, depending on the length of the emitted spin wave (or its oscillation frequency). As shown in Fig. 1a, the frequency changes (increases) with current, thus leading to the power oscillations as function of the current value.

These results clearly show that even the point contacts without any current can substantially affect the ST-induced oscillation dynamics of a multicontact system.

### B. Dependence of the system dynamics on the intercontact distance

General trends of the system dynamics and its synchronization behaviour with increasing the distance between nanocontacts are shown in Fig. 2. Here the color maps of the magnetization oscillation power $P(f, I_1)$ as the function of frequency $f$ and the current through the right contact $I_1$ are collected for various intercontact distances $\Delta L$ (defined on the scheme at the top of this figure). The current through the left contact $I_2 = 6.7$ mA is the same for all distances. First of all we note, that our simulations can not be considered as fully analogous to experimental measurements in the 'current ramping regime', because for each current value simulations have been performed separately: starting from the equilibrium magnetization configuration, both currents have been switched on (either instantaneously or linearly, see above) and magnetization oscillation trajectories have been recorded during 200 ns of the 'physical' simulation time. Afterwards, the power spectra of the time dependencies of the magnetization projections averaged over the areas under the point contacts were calculated for each value of $I_1$. Finally, spectra obtained this way at all currents $I_1$ for the given intercontact distance $\Delta L$ were combined into a 2D power spectral map. In Fig. 2 we maps for oscillations of the $m_x$-component are shown.

The first qualitatively important feature of the oscillation dynamics in our system is the absence of any synchronization when the contacts are placed *too close* to each other. On the upper left map in Fig. 2 we show an example of the corresponding behaviour for $\Delta L = 500$ nm. In the steady-state dynamical regime the system exhibits very broad spectral lines with the line width ~ 1 GHz, indicating the quasichaotic character of magnetization oscillations (we recall, that for a system with a *single* contact and the same nanostripe cross-section and material parameters we observe regular oscillations with the line width $\Delta f < 10$ MHz, i.e., at least two orders of magnitude smaller).

The most probable reason for this quasichaotic magnetization dynamics accompanied by a huge broadening of the linewidth in a system with relatively closely placed contacts (we observe this behaviour up to $\Delta L \approx 750$ nm) is the mutual disturbance of the magnetization configurations under each contact by the high-amplitude non-linear spin waves emitted from another contact. Due to the absence of the geometrically caused decay of the wave amplitude (in standard point contact systems based on extended 2D thin films an emitted wave decays as ~ $1/r$ even in the absence of the Gilbert damping), this mutual influence of the point contact oscillators destroys the regular character of ST-induced oscillations, thus making any synchronization impossible.

Beginning from the intercontact distance $\Delta L \approx 800$ nm, the synchronization stars to establish itself, leading to the clearly visible current synchronization region; corresponding example is shown in Fig. 2 for $\Delta L = 1000$ nm, where the synchronization is observed for 6.5 mA < $I_1$ < 6.9 mA. The mutual disturbance of the magnetization configurations caused by the waves emitted the contacts is still rather strong, so that the linewidth for this intercontact distance is about $\Delta f \sim 50$ MHz, but the presence of a synchronization is unambiguous. This picture qualitatively agrees with the experimental demonstration of the synchronization in Fig. 2a from [28] and the theoretical analysis of the mutual synchronization in two-contacts system shown in Fig. 2 from [23].

When the distance between the contacts is increased further, the linewidth becomes smaller, and starting from $\Delta L = 1500$ nm it is below our resolution limit $\Delta f_{min} \sim 10$ MHz, which value is determined by the maximal 'physical' simulated time. The current region where the synchronization is observed, first slightly increases, reaching its maximum (6.4 mA $< I_1 <$ 7.2 mA) for $\Delta L = 1500$ nm and then decreases with increasing $\Delta L$, until the synchronization disappears at $\Delta L = 3000$ nm.

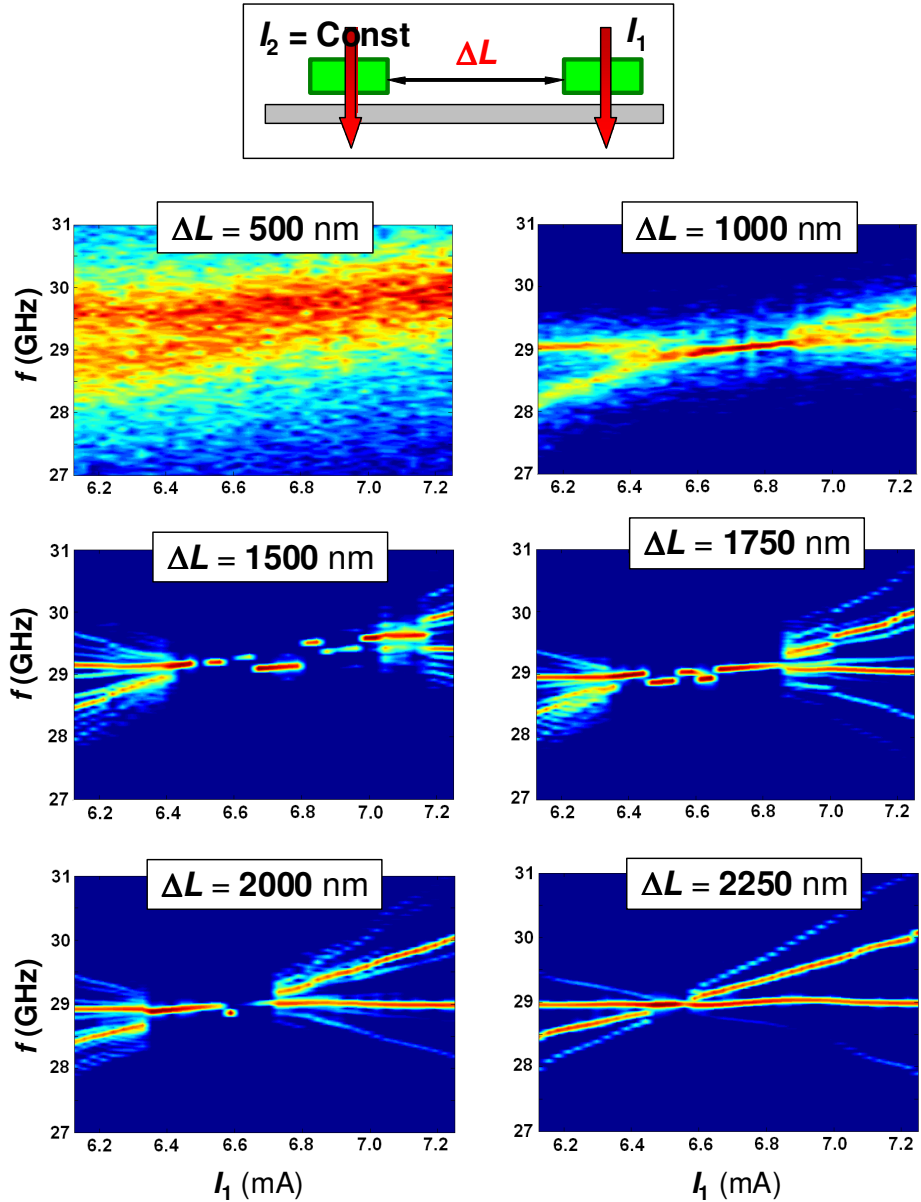

Fig. 2. Upper scheme: system geometry and current notation. 2D color plots: oscillation power $P_x(I_1, f)$ as function of $f$ and current strength in the 1$^{st}$ contact $I_1$ for various intercontact distances $\Delta L$ as shown in the legends (in all cases $I_2 =$ Const $= 6.7$ mA). Note very broad spectral lines (meaning quasichaotic magnetization dynamics and absence of any synchronization) for the smallest intercontact distances $\Delta L = 500$ nm. Synchronization can be observed starting from $\Delta L = 1000$ nm.

Here we would like to point out that the maximal distance for which we could observe the synchronization of STNOs on a quasi-1D nanostripe ($\Delta L_{max} = 3000$ nm) is *several times larger* than the maximal distances for which such a mutual synchronization was found up to now both experimentally and in simulations. In first successful synchronization experiments, the distance between nanocontacts was 500 nm [28] and 120 nm [29] (in the last paper, the maximal intercontact distance for a synchronization was estimated to be ≈ 200 nm). In [30], the synchronization of 4 nanocontacts placed in the corners of a square with the side $a$ = 500 nm was achieved, whereas for $a$ = 2000 nm oscillations of different contacts were independent on each other. Chen and Victora [26] have observed the STNOs synchronization simulating a system of two contacts placed also only maximal 500 nm apart.

As mentioned above, the principal difference between all these systems and our geometry is that in all cited papers experiments and/or simulations were performed on point contacts placed on top of extended 2D thin films. In such systems, the geometrically caused $1/r$-decay of the spin wave amplitude in 2D leads to much less favorable conditions for the synchronization of two STNOs, than in a quasi-1D nanowire studied in this report. In addition, thermal fluctuations present in real experiments performed at room temperature, may also reduce the maximal synchronization distance; however, it is unlikely that this effect can lead to the several times decrease of $\Delta L_{max}$ in quasi-1D waveguides. In principle, an analytical estimation of the intercontact coupling energy is needed in order to determine the relevance of the effect of thermal fluctuations for various intercontact distances (by comparing this coupling energy with the thermal energy). Unfortunately, synchronization in our system is most pronounced in the strongly supercritical regime - the relation $(I - I_{cr})/I_{cr}$, where $I_{cr}$ is the oscillation onset current, is not small - so that it is difficult to make such an estimation using the available analytical techniques (see, e.g., [23]).

### C. Existence of multiply synchronized states for the same current values through the contacts

Another non-trivial feature of the synchronization dynamics in the studied system is the existence of several possible steady precessional states in the synchronized regime. These multiply states can be seen very well in Fig. 2 especially for $\Delta L$ = 1500 nm and $\Delta L$ = 1750 nm: within the current region where the synchronization occurs the oscillation frequency as a function of current exhibits several relatively small but clearly discontinuous jumps.

To obtain an additional prove that our system possesses several possible synchronized states for the same system parameters (in particular, for one and the same current), we have performed two simulation series using two different protocols of the current increase as described in Sec. II (we remind that in the protocol A, we increased the currents in both contacts instantaneously, whereas in the protocol B, the current through the second contact $I_2$ was increased from zero to its final value *linearly* in time within 10 ns). Oscillation power maps shown in Fig. 3 for these two protocols are clearly different, thus demonstrating that for one and the same current values (within the synchronization region) multiply steady precessional states do exist. Using different current increase schemes, we have found up to 3 distinct synchronized states for one current value. Spatial maps of the oscillation phase (not shown) revealed, that magnetization oscillations under the two contacts can be either in-phase ($\Delta\phi = 0$) or out-of-phase ($\Delta\phi = \pi$) in these different synchronized states.

We point out, that this observation does not mean that jumps on the $f(I)$ dependence should occur in a real 'current ramping' experiment, because we simulate the magnetization dynamics for each current independently. In contrast to our simulations, in 'real' measurements the steady-state dynamical state at each given current value evolves out of the dynamical state established for the previous current. However, the existence of many precessional steady states suggests, that frequency jumps might be observed if transitions between these states occurr due to, e.g., thermal fluctuations in a real system.

When considered from the point of view of a general theory of non-linear coupled oscillators (see, e.g., [39]), the observed phenomenon confirms the existence of multiply attractors in our system. Thus the dependence of the final dynamical steady state on the current ramping protocol in our system is closely related to the synchronization dependence on the initial conditions for a system of serially connected STNOs predicted theoretically in [40, 41].

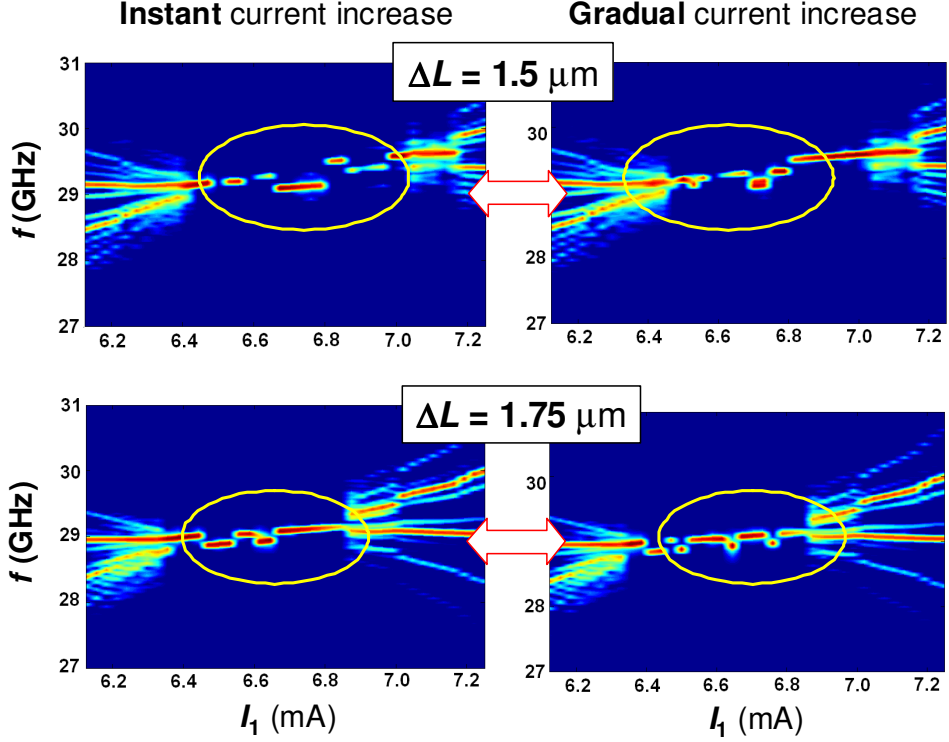

Fig. 3. Examples of spectral power maps $P_x(I_1, f)$ for two intercontact distances $\Delta L$. In all cases, the final value of the second current is $I_2 = 6.7$ mA. Left column: $I_2$ was increased from zero to its final value *instantly*. Right column: $I_2$ was increased *linearly* in time within $\Delta t = 10$ ns. Ellipses accentuate the areas of spectral maps where different synchronization behaviour for these two simulation protocols is clearly seen.

### D. Non-monotonous dependence of the synchronized oscillation power on the current value

As it can be seen already in Fig. 2 and 3, for the given intercontact distance the oscillation power depends non-monotonously on the current strength. To emphasize this complicated power vs current dependence, we show the $P_x(I)$-curves for several selected intercontact distances $\Delta L$ in Fig. 4. Three main reasons are responsible for this non-monotonous behaviour: (i) quasichaotic magnetization dynamics (for small intercontact distances), (ii) systematic change of the synchronized oscillation frequency with increasing current and (iii) jumps between various synchronization bands (states).

The first process – quasichaotic magnetization dynamics due to the mutual disturbance of the magnetization states within the contact areas by spin waves coming from another contact – leads to irregularities of the magnetization trajectories, resulting in a considerable broadening of spectral lines for intercontact distances up to $\Delta L = 1000$ nm. Due to these irregularities, averaging of the oscillation power over 200 ns physical time is not sufficient to fully suppress random power fluctuations, which are most pronounced in the upper left panel of Fig. 4, where $P_x(I)$ for $\Delta L = 500$ nm is drawn. With increasing simulation time these fluctuations should disappear; however, we did not study this process in more details, because this regime is not really interesting neither from the fundamental nor from the applied point of view.

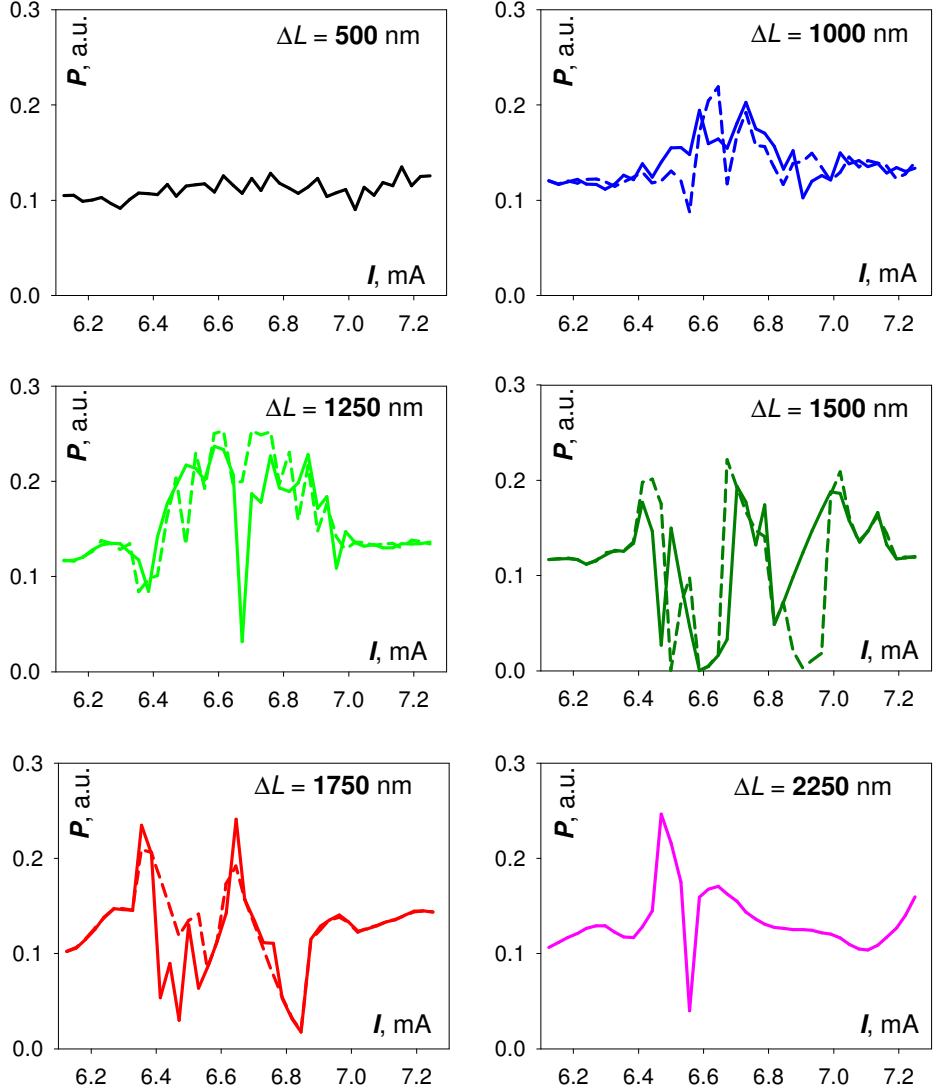

Fig. 4. Dependencies of the oscillation power $P_x$ on the current strength in the 1$^{st}$ contact $I_1$ for various intercontact distances $\Delta L$ indicated on the subplots (again, in all cases the final value of the 2$^{nd}$ current is $I_2 = 6.7$ mA). Solid lines: $P(I_1)$ for the case when $I_2$ was increased *instantly*. Dashed lines: $P(I_1)$ for $I_2$ increased to its final value *linearly* in time within $\Delta t = 10$ ns. For $\Delta L < 700$ nm only quasichaotic oscillations without synchronization are observed, for $\Delta L > 2200$ nm the same results are obtained for both simulation protocols.

The second reason is systematic change of the synchronized oscillation frequency when the current is increased, which means a systematic change of the spin wave length. Hence the interference of waves emitted from two contacts gradually changes its character from the constructive to the destructive one and back, resulting in regular power oscillations with increasing current. The same reason is responsible for power oscillations shown in Fig. 1 (we remind that the second wave is generated in this case ($I_2 = 0$) by the initial wave reflection from the non-homogeneity of the magnetization configuration created by the stray field of the 2$^{nd}$ CoFe nanosquare). Among the simulated systems, where the current flows through both contacts, analogous behaviour was observed in [26], where the interference of spin waves emitted from two contacts placed on the top of an extended thin film was studied.

The last reason for the non-monotonous $P_x(I)$-dependence in our simulations is the above mentioned existence of several possible steady precessional states. In Fig. 4, each panel contains two curves: the solid line for the instant increase of $I_1$ and the dashed one – for the linear increase of $I_1$ (see previous section for details). These two curves do not coincide for intermediate distances 1000 nm < $\Delta L$ < 2000 nm, demonstrating once more than the steady precessional state achieved by the system strongly depends on its history due to the presence of multiply dynamic fixed points.

## IV. CONCLUSION

Summarizing, we have presented systematic numerical studies of the magnetization dynamics induced in a thin NiFe nanostripe by a spin-polarized current injected via square-shaped CoFe nanomagnet(s) in the CPP-geometry. In this quasi-1D system, magnetization oscillations induced by a spin torque in the nanostripe areas under the CoFe point contacts may be synchronized (phased-locked) due to the intercontact interaction mediated by spin waves emitted out of the contact areas. An oscillatory mode *propagating* along the nanowire does exist in this system when a sufficiently strong *out-of-plane* field is applied. This mode is a highly promising candidate for obtaining a synchronization of point contact nanooscillators, because in a quasi-1D waveguide the geometrically caused decay of the spin wave amplitude (inevitably present in 2D and 3D systems due to the wave front expansion) is absent.

On one hand, our numerical results confirm this expectation: we have demonstrated that the synchronization of two STNOs on such a nanowire is possible up to the maximal distance between the contacts $\Delta L$ = 3 mkm, when the nanowire is only 50 nm wide. On the other hand, studying the synchronization in more details, we have found that the achievement of a reliable synchronization is much more complicated than could be expected based only on the dimensional argument presented above. In particular, we have shown that there exist a *minimal* distance between the nanocontacts, below which such a synchronization can not be obtained due to a strong mutual disturbance of the magnetization configuration under each contact by the spin wave emitted from the other contact. Further, we have demonstrated that for the given distance between the contacts, the power of synchronized oscillations depends *non-monotonously* on the current value in one of the contacts, when current value in another contact is constant (due to the wave interference). The last important finding is the existence of several synchronized states for one and same current values through the contacts, so that the final steady precessional state depends on initial conditions. All these features make the achievement of a stable synchronization of two and more STNOs to a highly non-trivial task, requiring a careful optimization of the system design.